\documentclass[prb,aps,twocolumn,showpacs,preprintnumbers,amsmath,amssymb,superscriptaddress]{revtex4-1}

\usepackage{graphicx}
\usepackage{dcolumn}
\usepackage{color}

\begin{document}


\title{The importance of level statistics for the decoherence of a central spin due to a spin environment}

\author{H{\aa}kon Brox}%
\affiliation{Department of Physics, University of Oslo, PO Box 1048 Blindern, 0316 Oslo, Norway}

\author{Joakim Bergli}%
\affiliation{Department of Physics, University of Oslo, PO Box 1048 Blindern, 0316 Oslo, Norway}

\author{Yuri M. Galperin}%
\affiliation{Department of Physics, University of Oslo, PO Box 1048 Blindern, 0316 Oslo, Norway}
\affiliation{Centre for Advanced Study, Drammensveien 78, Oslo, Norway 0271, Oslo, Norway}
\affiliation{A. F. Ioffe Physico-Technical Institute of Russian Academy of Sciences, 194021 St. Petersburg, Russia}

\date{\today}

\begin{abstract}
We study the decoherence of a central spin-$1/2$ due to a closed environment
composed of spin-$1/2$ particles. It is known that a frustrated spin
environment, such as a spin glass, is much more efficient 
for decoherence of the central spin than a
similar size environment without frustration. We construct a Hamiltonian
where the degree of frustration is parametrized by a single parameter $\kappa$.
By use of this model we find that the environment can be classified by two distinct regimes with respect to
the strength of level repulsion. These regimes behave qualitatively
different with respect to decoherence of the central spin and might
explain the strong enhancement of decoherence observed for frustrated
environments.

\end{abstract}

\maketitle

\newcommand{\eq}{\! = \!}
\newcommand{\keq}{\!\! = \!\!}
\newcommand{\kadd}{\! + \!}

\section{Introduction}
Quantum decoherence, where coherence in a quantum system is reduced
due to interaction with its environment is a fundamental concept of
physics.  Testing of theories that go beyond unitary quantum
mechanics~\cite{leggett,ghirardi,wezelprl} requires deep understanding
and control of the decoherence process in order to distinguish the
breakdown of unitarity predicted in these theories from decoherence.
Decoherence is also a fundamental problem in the branch of
nanoscience, where one seeks to use and manipulate quantum states for
application. Coherent manipulation and storage of quantum information
are required in order to construct a working quantum computer and rely
on reducing decohering interaction between its basic elements, the
qubits, and their environment.

Recently there has been increased experimental interest in electronic
spin systems, where the most prominent source of decoherence is
thought to be electronic, or, in samples with high purity, nuclear
spins. These systems are nitrogen-vacancy centers in
diamond,~\cite{gaebel_nphys,Hanson_science} semiconductor quantum
dots~\cite{ladd,hanson_rmp,coish,tsyplyatyev} and large-spin magnetic
molecules.~\cite{takahashi,stamp_nature} In addition, fluctuating two
level defects are thought to be the major source of decoherence in
solid state Josephson junction qubits, see Ref.~\onlinecite{bergli} for a review.
  The coherence of
a single spin interacting with a spin bath has been studied
extensively in the limit of a non-interacting
bath.~\cite{stamp} Decoherence due to interacting spins have also been
studied recently in the weak coupling limit~\cite{camalet_prb} and it
was found that the coherence of the central spin decays rapidly when
the environment is close to a phase transition.~\cite{camalet_prl}

Decoherence, relaxation and thermalization of a central system coupled
to a closed, finite-size spin bath environment has been investigated
in Refs.~\onlinecite{yuan_jetp06,yuan_prb08,yuan_pra07,lages_pre05,melikidze_prb04,yuan_jpsj09}.
In Refs.~\onlinecite{yuan_jetp06,yuan_prb08}, decoherence of a
two-spin system was studied, and a large enhancement of decoherence
was found for frustrated spin environments, the main conclusion being:
  \emph{``For the models
under consideration, the efficiency of the decoherence decreases
drastically in the following order: Spin glass - frustrated
antiferromagnet - Bipartite antiferromagnet - One dimensional ring
with nearest neigbour antiferromagnetic
interactions''}.\cite{yuan_jetp06} A similar study found that the same
was true also with regards to relaxation towards the ground state of
the central system. Namely,  frustrated environments are more efficient in
relaxing the central system compared to an ordered environment.\cite{yuan_pra07}
  Whether or not the bath is chaotic or integrable
was shown to be of less importance.\cite{lages_pre05} Frustrated spin
systems have been suggested to exist as localized electron states on 
the surface of superconducting quantum interference devices, 
such as SQUIDs and flux qubits,\cite{clarke_09}
where they are thought to be a major source of magnetic flux noise.

However, a detailed understanding of the physics behind the importance
of a frustrated environment is still lacking. In this work we construct a
model where we can continuously tune the degree of frustration in the
environment by a single parameter $\kappa$, confirming that frustrated environments reduce the
coherence of the central spin much more efficient compared to an
environment with low degree of frustration, as previously found in
Refs.~\onlinecite{yuan_jetp06,yuan_prb08}. Using this model we study
the structure of the eigenvalues of Hamiltonian  of the
environment, $H_E$, in the presence of a central spin.

We find that we can explain the mechanism behind the efficiency of the
frustrated environment by the structure of the eigenvalues of
$H_E$. The frustrated environment can be characterized by a
Wigner-like distribution of eigenvalues, and therefore has large
repulsion between energy levels. The presence of an external object,
like a central qubit, will therefore result in the mixing of a large
fraction of the eigenvectors of the unperturbed system. In an ordered
environment, however, the level repulsion is very weak, and coupling to
the central spin will only alter the set of eigenvectors of the
environment slightly, preserving the coherence of the central
spin. 

The link between the response of the eigenvectors of $H_E$ to an
external perturbation, and the decoherence of a central spin 
is found as follows. 
The initial state of the complete system is
\begin{equation}\label{phi0}
|\Phi(t=0)\rangle=(1/\sqrt{2})\left(\left|\uparrow\right\rangle_S
  +\left|\downarrow\right\rangle_S\right)\otimes\left|\psi_0\right\rangle_E,
\end{equation}
where the subscripts $S$ and $E$ denote the central system and the
environment, respectively, and we have for simplicity assumed the
central system to be in an initial symmetric superposition. The state
$\left|\uparrow\right\rangle$ means that the system is in the
eigenstate of the operator $S^z$ with eigenvalue $+1/2$ and
$\left|\psi_0\right\rangle_E$ is the initial state of the environment.
If the system and environment are coupled, the state of the system
influences the dynamic evolution of the environment, and we can write
the linear time evolution of the composite system as 
\begin{equation}\label{timeEv}
\left(\left|\uparrow\right\rangle_S
   +\left|\downarrow\right\rangle_S\right)\left|\psi_0\right\rangle_E
 \to \left|\uparrow\right\rangle_S\left|\psi(t)^{\uparrow}\right\rangle_E+\left|\downarrow\right\rangle_S\left|\psi(t)^{\downarrow}\right\rangle_E
\end{equation}
where $|\psi(t)^{\uparrow}\rangle_E$ denotes the time evolution of the
environment conditioned upon that the initial state of the central
system is $\left|\uparrow\right\rangle_S$. For now we assume that the system-environment
coupling, $H_{SE}$, commutes with $S^z$, so that transitions between
the levels of the central system is prohibited.  
We will characterize the decoherence by the off-diagonal matrix element of the density matrix,
\begin{align}
\rho_{\uparrow\downarrow}^S=\langle\psi(t)^{\downarrow}|\psi(t)^{\uparrow}\rangle.
\end{align}
Let us expand the state of the environment in the set of eigenstates, 
\begin{align}
|\psi(t)^{\uparrow}\rangle
  =\sum\limits_n\langle n^{\uparrow}|\psi_0\rangle|n^{\uparrow}\rangle 
  e^{iE_n^{\uparrow}t},
\end{align}
where $|n^{\uparrow}\rangle$, $E_n^{\uparrow}$ denote the eigenstates
and eigenvalues of the environment conditioned upon that the central
spin points up, and similarly in the case where the central spin
points down (throughout the paper we put $\hbar=1$).  Then the time evolution of the off diagonal element of the
density matrix is given by the expression
\begin{equation} \label{olap}
\rho_{\uparrow\downarrow}^S(t) 
=\sum\limits_{n,m}\langle n^{\uparrow}|\psi_0\rangle\langle\psi_0|n^{\downarrow}\rangle\langle n^{\uparrow}|m^{\downarrow}\rangle e^{i(E_n^{\uparrow}-E_m^{\downarrow})t}.
\end{equation}
Thus $\rho_{\uparrow\downarrow}^S$ is determined 
by the magnitude of the overlap elements, unless the levels are degenerate.
For degenerate states
the corresponding phase factors of the overlap with
each eigenstate of the degenerate level oscillate with the same phase.

The analysis is simplified if we assume that only the upper state of
the central system couples to the environment, and that the environment
is prepared in its ground state.  In this case Eq.~(\ref{olap})
simplifies to
\begin{equation}
\rho_{\uparrow\downarrow}^S (t) =\sum\limits_{n}|\langle n^{\uparrow}|0\rangle|^2 e^{i(E_n^{\uparrow}-E_0)t}
\label{olapsimp}
\end{equation} 
and the picture is more transparent.  

Evolution of $\rho_{\uparrow\downarrow}^S $ is then determined by quantum beatings 
between the overlap contributions oscillating at frequencies $(E_n^\uparrow -E_0)$,
i.~e., by the differences between
eigenvalues of $H_E$ and the eigenvalues of the environment in
presence of the central spin.  
In the following
we will investigate this further by numerical study of an explicit model.

The paper is organized as follows. In Sec.~\ref{model} we describe our model of a
central spin-$1/2$ interacting with a spin environment with tuneable degree of frustration.
In Sec.~\ref{sec:overlap} we study the different regimes
of decoherence of our model, while in Sec.~\ref{levels} we explain the
physical mechanism behind the enhancement of decoherence by frustration in detail.
Furthermore, in Sec.~\ref{temp} we describe the sensitivity to the initial state and
in Sec.~\ref{magnetic} we suggest a method to reduce the negative impact from frustrated environments
on coherence.
Finally the results will be discussed in Sec.~\ref{discussion} and we conclude in Sec.~\ref{conclusion}.

\section{Model}
\label{model}

We model a central spin-$1/2$ interacting with a spin environment by the Hamiltonian 
\begin{align}\label{ham}
H&=H_S+H_{SE}+H_E,\\
H_{SE}&=\frac{1}{2}\sum\limits_{i}\Delta_{i}\left(S^z-\frac{1}{2}\right)s_i^{z} ,\nonumber\\
H_{E}&=\sum\limits_{i,j,\alpha}\Omega_{ij}^{'\alpha}s_i^{\alpha}s_j^{\alpha} \nonumber
\end{align}
where $H_{S}$, $H_{SE}$ and $H_E$ are the Hamiltonians for the central
spin, the spin-environment coupling and the environment,
respectively, $S$ is the operator of the central spin, while $s_i$ are the 
operators of the environmental spins. We set both the energy splitting and the tunneling element
 of the central system to zero.
The parameters $\Delta_{i}$ and
$\Omega_{ij}^{'\alpha}$ specify the coupling strength along the
$\alpha$-axis between the central spin and the environment and the
intraenvironment coupling, respectively. The  parameters $\Delta_{i}$ are chosen
randomly in the interval $[-\Delta,\Delta]$.

In order to study the importance of frustration we specify $H_E$ as
\begin{equation} \label{He}
H_{E}=-\Gamma\sum\limits_{i,j,\alpha}\left[(1-\kappa)s_i^zs_j^z
  +\kappa\Omega_{ij}s_i^{\alpha}s_j^{\alpha}\right]
\end{equation}
where $\Omega_{ij}$ is a random number in the interval
$[-\Omega,\Omega]$.
The degree of disorder is then parametrized by $\kappa\in[0,1]$. 
In this model we can continuously tune our environment by the parameter
$\kappa$ from a perfect ferromagnet ($\kappa=0$) to a highly
frustrated spin-glass ($\kappa=1$).
In the following all the energies will be measured in the units of $\Gamma$, 
therefore $\Gamma =1$. Correspondingly, time is measured in units of $\Gamma^{-1}$.

The simulation procedure is the following.  We select a set of model
parameters. Then we compute the eigenstates and eigenvalues of $H$ 
by numerical diagonalization.  
The composite system is prepared in the
state \eqref{phi0}
where $|\psi_0\rangle$ is the initial state of the environment and the
central spin is prepared in a superposition of eigenstates of $S^z$. 
Unless otherwise stated, the initial state of the environment is always the
ground state in the absence of coupling,
$|\psi_0\rangle_E=|0\rangle_E$. In general, the initial state is
therefore a complicated superposition of eigenstates of the composite
system $H$.

Decoherence is in this model solely due to entanglement between the
central system and the environment. The
state evolves according to the Schr\"{o}dinger equation into an, in
general, entangled state as in Eq.~\eqref{timeEv}.
The reduced density matrix of the system is obtained by tracing over
the degrees of freedom of the environment: $\rho^S(t)=\mathrm{Tr}_E\{\Phi(t)\}$.

\section{Results}
\label{results}
\begin{figure}[t]
\centerline{
  \includegraphics[width=8cm]{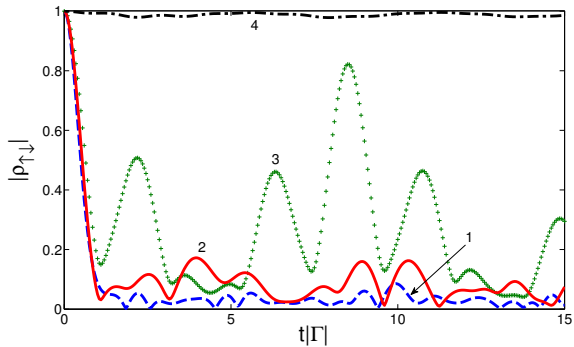}
}
  \caption{(Color online) Time evolution of the off-diagonal density matrix element,
 $|\rho_{\uparrow\downarrow}^S(t)|$, for $N=9$ spins in the
 environment and different values for the disorder parameter $\kappa$,
 ranging from: $1.$ the spin glass phase ($\kappa=1.0$, dashed, blue)
 to $4.$ the ferromagnetic phase ($\kappa=0.1$, dashed/dots, black). 
 Other arrangements are frustrated ferromagnet $\kappa=0.5$ ($2$, solid, red) and 
 for comparison we plot the time evolution for the completely disconnected
 bath with Heisenberg like $H_{SE}$ and $\kappa=1$, $\Omega=0.0$ ($3$, `+', green),
in this configuration there is no correlations between the different systems in the environment.
The strength of the system-environment coupling is $\Delta=3.0$.
}
\label{timeevo_sqr_500}
\end{figure}

Using the simulation procedure described above we can study the
dynamics of the reduced density matrix of the central system. The time
evolution of the off-diagonal element of the density matrix for
different values of the environment parameters, is shown 
in Fig.~\ref{timeevo_sqr_500}. We find that in general, higher degree of frustration, 
controlled by the parameter $\kappa$, result in stronger and more robust 
decay of  $\rho_{\uparrow\downarrow}^S$.
The initial evolution
is similar and Gaussian in time  for all values of $\kappa$, however for smaller $\kappa$ we find rapid
revivals of coherence in the central system. 

From Fig.~\ref{timeevo_sqr_500} we see that it is useful to
distinguish between the initial decoherence and the efficiency of
decoherence. We define the \textit{initial decoherence} as the evolution of coherence 
in the central system in the characteristic time during which $\rho_{\uparrow\downarrow}^S$ decays by a factor $e$
and the \textit{efficiency of decoherence} as the mean value of the off-diagonal elements of the
density matrix over a period that is large compared with the dynamics
of the environment. 

From Fig.~\ref{timeevo_sqr_500} we thus find that
initially $|\rho_{\uparrow\downarrow}^S|$ decays following the Gaussian law, 
$\rho_{\uparrow\downarrow}^S \propto e^{-(t/t^*)^2}$, 
with practically $\kappa$-independent decay time $t^*$. 
The efficiency of the decoherence is, however, much higher for the
frustrated environment $\kappa=1.0$. If the efficiency of decoherence
is low, as for the ferromagnetic environment, the error might be
corrected by use of quantum error correction.~\cite{chang} In fact, we
show in Fig.~\ref{timeevo_sqr_500} that a completely disconnected bath,
$\Gamma=0$, gives stronger decoherence than the ferromagnetic bath.

The picture we obtain is the following. The decoherence of the central
spin is dependent on the sensitivity of the environment to the state
of the central system. The response of the environment to an external system is closely related to the
sensitivity  of the Hamiltonian of the environment to a small perturbation. 
The latter can be, in turn, related to the so-called  Loschmidt echo defined as the overlap
 between the two states evolving  from the same initial wave function under the influence of two
distinct Hamiltonians, the unperturbed $H_0$ and a perturbed one
$H_\Lambda=H_0+\Lambda$, see Ref.~\onlinecite{zurek_prl_03} for
details. Therefore, in most cases, the Loschmidt echo of the environment and
the efficiency of the decay of the off-diagonal elements of $\rho_S$
will be strongly correlated, even though there are
exceptions.~\cite{casabone_epl_10}  Thus our analysis applies to the
purity of the central system as well as to the sensitivity to
perturbations of the environment.
\begin{figure}[htb]
\centerline{
  \includegraphics[width=8cm]{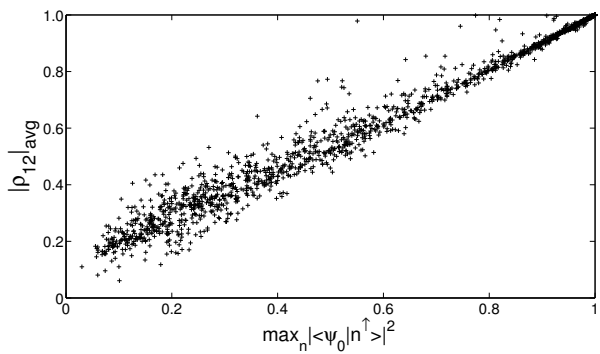}
}
  \caption{Correlation between
$|\rho_{\uparrow\downarrow}^S(t)|_{\text{avg}}$, and the largest overlap
element between the ground state of $H_E$ and the set of eigenstates
$\{|n^{\uparrow}\rangle\}$ of the environment in presence of $H_{SE}$.
We see that the size of the largest overlap element is strongly
correlated with the decoherence,
$|\rho_{\uparrow\downarrow}^S(t)|_{\text{avg}}$, which  is defined by the average of
$\rho_{12}(t)$ over the interval $t\in[200,300]$, i.~e. after the
initial rapid decoherence has taken place. We call this the
efficiency of decoherence.\cite{yuan_pra07} The details of
averaging do not matter as long as $t$ is much larger than the
initial decay time, and the averaging interval is much larger than the
correlation length of the oscillations. The statistics are obtained by
sampling over the parameter range $\Omega\in[0,1]$,
$\Delta\in[0,3]$. The number of spins in the environment is
$7$.}
\label{corrdec}
\end{figure}

The sensitivity of the state of the environment to a
perturbation (in our case, to a flip of  the central spin) and, therefore, 
the efficiency of decoherence can be characterized by overlaps
 between the initial state of the environment,
$\left|0\right\rangle_E$, and the set of eigenstates of the
environment in the presence of the perturbation,
$\{\left|n^\uparrow\right\rangle\}$. 
We find that the largest of the overlap elements serves as a very good 
indicator for the decoherence of the
central spin. We measure the efficiency of the environment by
the modulus of the off-diagonal element of the reduced density matrix,
$|\rho_{12}|_{\text{avg}}$, averaged over the interval $t\in[200,300]$ that is long
compared to the typical oscillation periods in
$|\rho_{\uparrow\downarrow}^S (t)|$, cf. with  Fig.~\ref{timeevo_sqr_500}. 
The relationship between $|\rho_{12}|_{\text{avg}}$ and the largest overlap element is plotted in
Fig.~\ref{corrdec}. The fact that the largest overlap element
correlates so well with the decoherence suggests that the probability of finding
degenerate eigenstates among the states with the largest overlap element is relatively small
and that the detailed distribution of overlapping vectors $\{\left|n^{\uparrow}\right\rangle\}$
is less important.

In the rest of the article we will use numerical simulations to
clarify the difference with respect to decoherence of a central system
interacting with a ferromagnetic or a frustrated environment. In view
of the strong correlation demonstrated in Fig.~\ref{corrdec} we will use 
$\max_n|\langle n^\uparrow|0\rangle_E|^2$ as a measure of the
efficiency of the decoherence. 

\subsection{Decoherence in terms of the overlap with the initial state}
\label{sec:overlap}

We decompose the initial state of the environment in the eigenstates
of $H_E$ and use the ground state $\left|\psi_0\right\rangle_E=\left|0\right\rangle_E$ as
the initial state. In Ref.~\onlinecite{yuan_prb08}, decoherence was
studied both using the ground state as initial state and a random
superposition of eigenstates corresponding to ``infinite
temperature''. We will focus first on the ground state and address a more
complicated initial state in Sec.~\ref{temp}. In the absence of disorder,
$\kappa=0$, the ground state of $H_E$ will be the ferromagnetic state
where all spins point in the same direction
$\left|\uparrow\uparrow...\uparrow\right\rangle_E$, or in general a linear combination of
the two degenerate ground states. In order to avoid the exact
degeneracy, we use a small static symmetry breaking field acting on a
single spin in the environment.
\begin{figure}[h]
\centerline{
  \includegraphics[width=8cm]{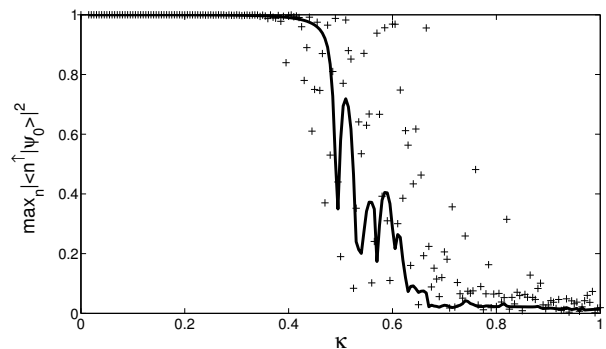}
}
  \caption{The largest overlap element plotted versus the disorder strength $\kappa$.
For small values of $\kappa$ the ferromagnetic ground state is energetically strongly favored and the perturbation 
represented by the central spin is not able to significantly alter the ground state. Close to $\kappa=0.5$ we find a 
``phase transition'' to a more disordered state. In this regime the coupling to the central spin is sufficient to alter the ground state of the environment. For $\kappa\approx 1.0$ the set of eigenstates are completely altered in the presence of the central spin, and the overlap with the original set is typically very small. The number of environmental spins is $N=9$, $\Delta=3.0$ and the same seed is used in generating the distributions of $\Omega_{ij}$ and $\Delta_{i}$ for each value of $\kappa$ (solid line), while the markers correspond to a random seed for each value of $\kappa$.} 
\label{N9P6_D}
\end{figure}

In the ferromagnetic phase, if the strength of the system-environment
coupling is weaker than the intra-environment coupling, $\Delta\ll
N\Omega$, the presence of the central spin will not alter the ground
state significantly. Therefore, the overlap between the ground state
of the isolated environment with the ground state of the perturbed
environment, $\langle 0|0^{\uparrow}\rangle_E$, will be very close to one
(i.~e., the magnitude of all the terms of Eq.\eqref{olapsimp} will be
close to zero except for the term $\langle 0|0^{\uparrow}\rangle_E$, where
$|0^{\uparrow}\rangle_E$ is the ground state of the environment given the
perturbation). Thus the ground state will still be ferromagnetic in
the presence of the central spin, which will therefore not entangle
sufficiently with its environment, preserving the coherence. In
Fig.~\ref{N9P6_D} we show numerical simulations for different values
of frustration in the environment. As long as the disorder parameter
$\kappa$ is small, the largest overlap element between the unperturbed
ground state of the environment $|0\rangle_E$ and the set of
eigenstates $\{|n^{\uparrow}\rangle\}$ when the system-environment coupling
$H_{SE}$ is turned on, is very close to one. The ground state is
ferromagnetic and the interaction with the central spin is
insufficient to break the ferromagnetic order.

If we now increase the disorder parameter $\kappa$, the ground state
of the environment will be only slightly altered, until the
frustration in $H_E$ given by $\kappa$, together with the frustrated
Ising type system-environment coupling $H_{SE}$, becomes large enough
to break the ferromagnetic order. This ``phase transition'' is evident
from Fig.~\ref{N9P6_D}, where in these particular simulations it takes place
at about $\kappa\approx 0.5$, but the value is in general dependent on
the size of the system, and the strength and nature of $H_{SE}$. The physics
during and after the phase transition will be addressed in more
detail in Sec.~\ref{levels}. 
We can make a rough estimate as follows: A single spin is in general
subject to two competing interaction effects, the ferromagnetic
interaction $(1-\kappa)\Omega N$ and the spinglass interaction
$\propto\kappa$. Assuming that the latter is random it should be of
magnitude $\kappa\Omega\sqrt{N}+\Delta$. The transition between
the ferromagnetic and the spinglass phase should therefore take
place at
\begin{align}
\kappa\Omega\sqrt{N}+\Delta\approx(1-\kappa)\Omega N.
\label{transition}
\end{align} 
If we insert the parameters from Fig. \ref{N9P6_D} we find the
critical value $\kappa=0.5$. 
Summarized, if the total frustration induced together by $\kappa$ and
$H_{SE}$ is insufficient to break the ordered ground state, both
$\left|0\right\rangle_E$ and $\left|0^{\uparrow}\right\rangle_E$ will have a large overlap
with one of the states $\left|\uparrow\uparrow...\uparrow\right\rangle_E$ or
$\left|\downarrow\downarrow...\downarrow\right\rangle_E$, according to
Eq. \eqref{olapsimp}, and in this regime the coherence of the central
system will be preserved.
\begin{figure}[htb]
\centerline{
  \includegraphics[width=8cm]{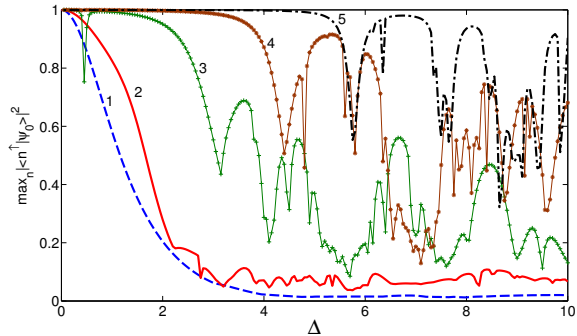}
}
  \caption{(Color online) The largest overlap element computed according to
  Eq.~\eqref{olapsimp}, plotted versus the perturbation strenght $\Delta$,
  for different values of the disorder parameter $\kappa$. The numbered
  curves correspond to the following values of $\kappa$: $1.$ spin glass
  $\kappa=1.0$ (dashed, blue), $2.$ $\kappa=0.8$ (solid, red), $3.$ $\kappa=0.6$ (`+', green), 
  $4.$ $\kappa=0.4$ (`$\ast$', brown), $5.$ $\kappa=0.2$ (dash-dotted, black). The environment is
  prepared in the ground state of $H_E$. The number of environmental
  spins are $N=9$, and the same seed is used for each value of
  $\Delta$. }
\label{multi}
\end{figure}

In Fig.~\ref{multi} we follow the largest overlap element
$\max_n|\langle n^{\uparrow}|0\rangle_E|^2$ as a function of
the strength of the system-environment coupling $\Delta$, keeping
$\kappa$ constant. In each of the simulations $H_{SE}$ is random and
Ising-like.  We find that for small values of $\kappa$, the strength
of the random, frustrated system-environment coupling $H_{SE}$ needs to
be sufficiently large in order to break the ferromagnetic interaction, in
accordance with Eq.~\eqref{transition}. 
Indeed, using Eq.~\eqref{transition} we predict the following values
for the critical $\Delta$
\[
\begin{array}{|c|c|c|c|c|c|}\hline\kappa&0.2&0.4&0.6&0.8&1\\
\hline
\Delta&  6.6 &  4.2 &   1.8  &  -0.6&    -3.0\\
\hline
\end{array}
\]
which agrees surprisingly well with Fig.~\ref{multi}.
Until ferromagnetic order is
destroyed by $H_{SE}$ the ground state of the perturbed system is very
close to parallell with the unperturbed ground state, $\langle
0^{\uparrow}|0\rangle_E\approx 1$. For larger values of $\Delta$, we
find strong oscillations in the size of the overlap element as a
function of $\Delta$. This effect does not take place for $\kappa\geq
0.8$.  In this regime the oscillations are less pronounced and the
decay of the overlap element takes place for smaller values of
$\Delta$. This regime is characterized by a highly frustrated ground
state in the absence of $H_{SE}$. The presence of the central spin
only alters microscopic details of the ground state, not its qualitative features.

In summary, an environment with frustrated interactions induces more
effective decoherence than an unfrustrated environment. This effect
can be quantified by the strength of a perturbation (here
$H_{SE}$) which alters the set of eigenstates $\{|n\rangle_E\}$. If the
environment is dominated by frustrated interactions the set of
eigenstates $\{|n^{\uparrow}\rangle_E\}$ in presence of the
perturbation $H_{SE}$ will in general be very different from
$\{|n\rangle_E\}$. We can think of this process as follows: In an
environment with a large number of opposing interactions and a large
set of almost degenerate low energy states, the presence of a central
spin will in general cause a rotation of a subset of the eigenvectors
$\{|n\rangle_E\}$. If there is a rotation and given that the subset
contains the ground state $|0\rangle_E$, the maximal overlap element
$\max_n|\langle n^{\uparrow}|0\rangle_E|^2$ and therefore the
coherence of the central spin will decay. We will discuss the detailed
physics behind this process in more detail in  Sec.~\ref{levels}.

\subsection{Decoherence in terms of avoided level crossings}
\label{levels}

In order to gain a deeper understanding of the differences between the
ordered and the frustrated environment with respect to dephasing of
the central spin, we study in detail the behavior of the eigenvalues.
We use the same model as defined previously by Eq.~\eqref{He} and
an Ising-like random $H_{SE}$.
Then we perform simulations where we gradually increase the coupling parameter $\Delta$ for different values of the
disorder parameter $\kappa$.

\begin{figure}[ht] 
\centering
%
\includegraphics[width=0.99\columnwidth]{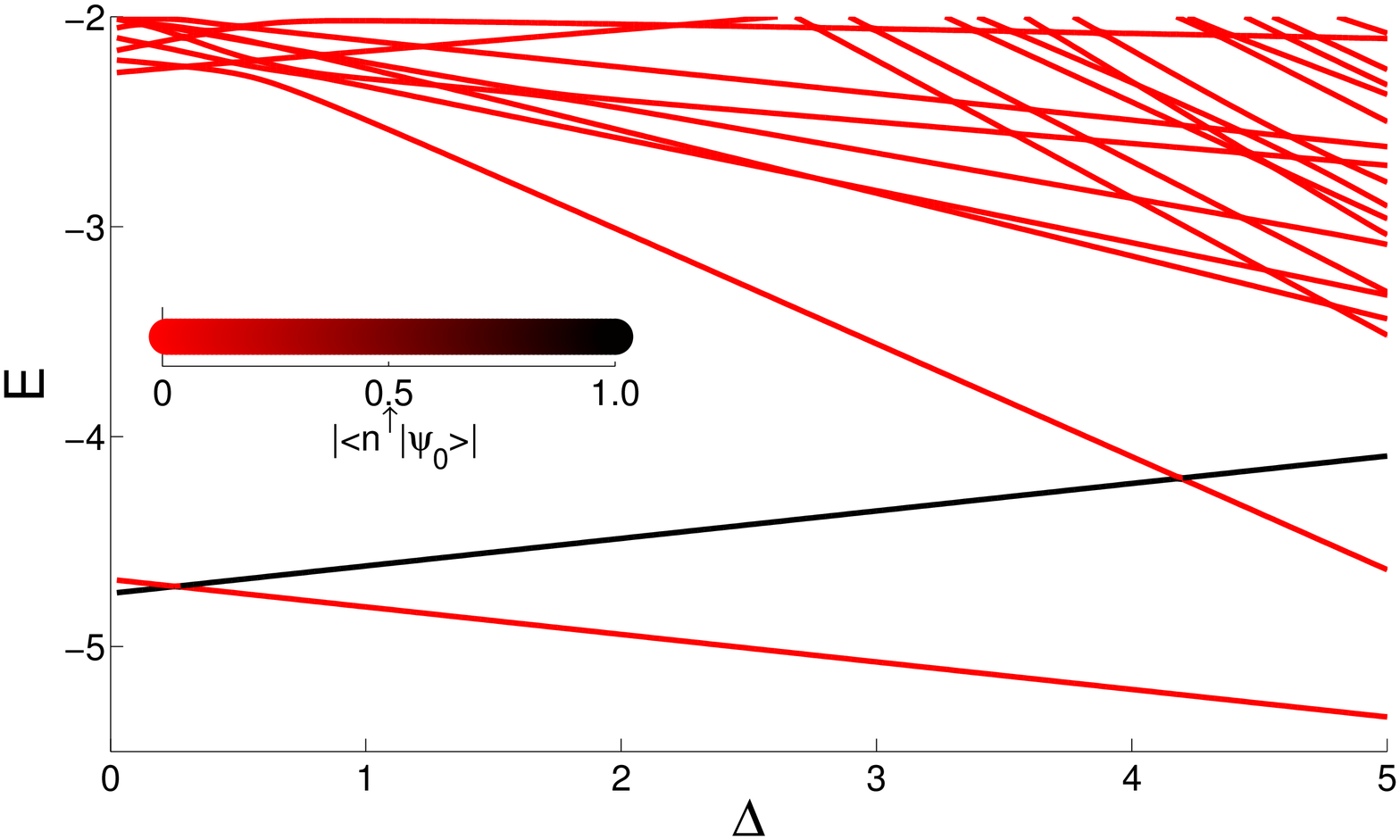}  \\ 
\includegraphics[width=0.99\columnwidth]{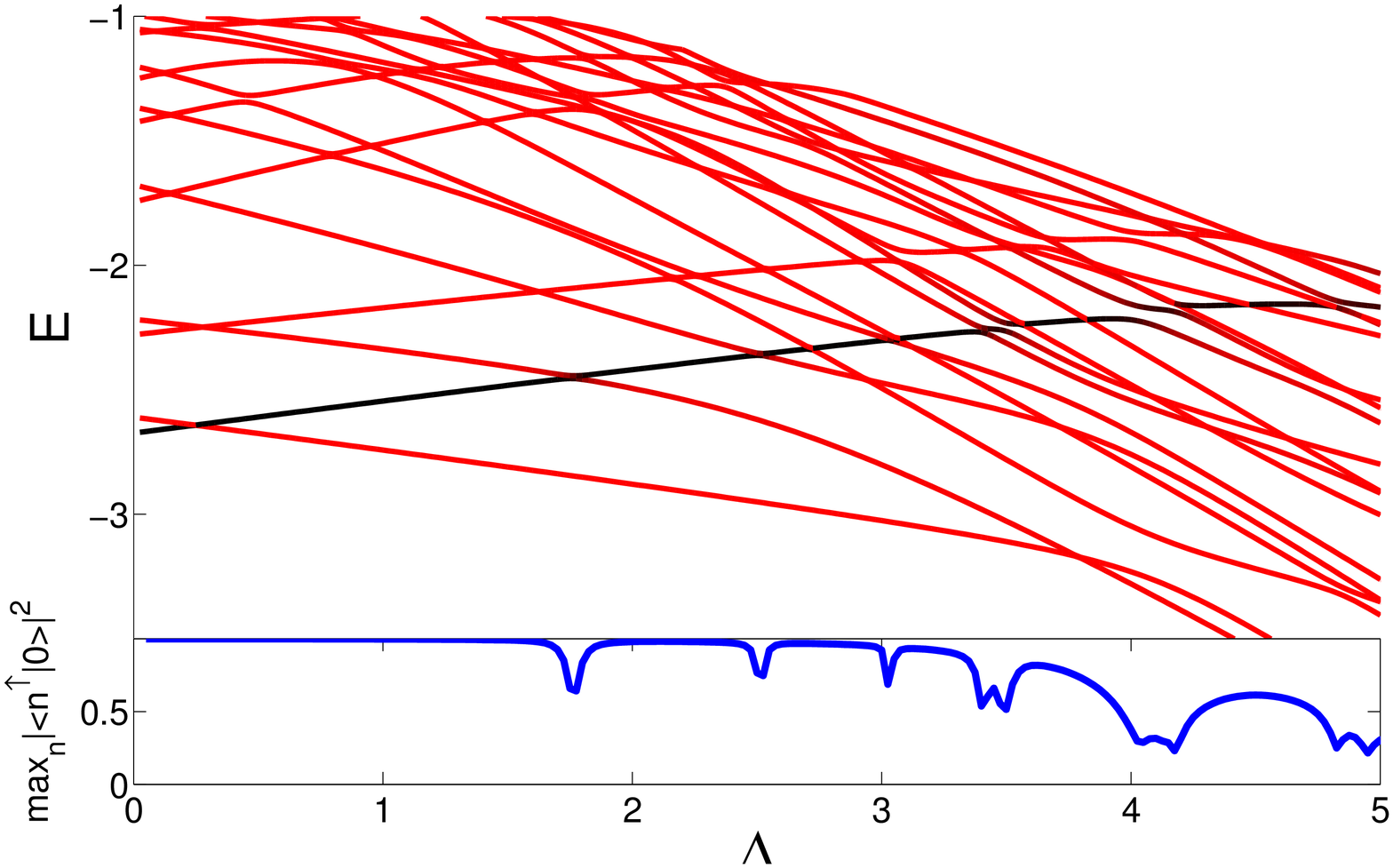} \\
\includegraphics[width=0.99\columnwidth]{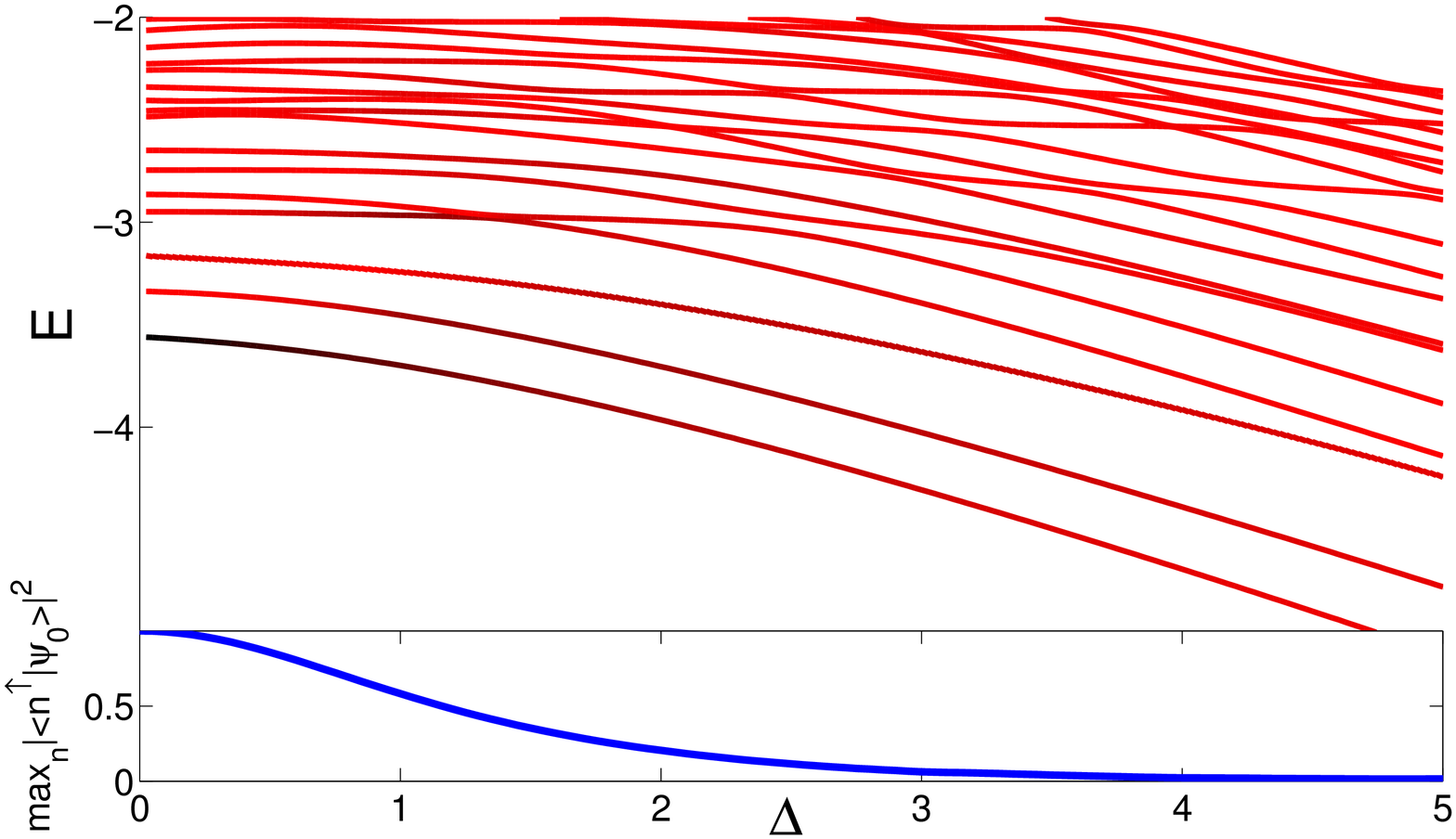} 
  \caption{(Color online) The $20$ lowest eigenvalues plotted against the perturbation strength $\Delta$ for different values of frustration in the environment. $\kappa =0.1$~(top), 0.5~(middle), 1.0 (bottom).
%
The overlap with the ground state of the unperturbed Hamiltonian $H_E$ is indicated by the color tone.
A large overlap element increase the darkness of the corresponding eigenvalue (color bar is shown in the upper figure).
The bottom plot shows the largest overlap element between the ground state of $H_E$ and the eigenstates of $H_E$ in presence of the interaction $H_{SE}$, $\max_n|\langle n'|0\rangle|^2$.
The number of spins in the environment is $N=8$. \label{ordered05}
} 
\end{figure}

In Fig.~\ref{ordered05}~(top), we plot the $20$ lowest eigenvalues against
the coupling strength $\Delta$.  The disorder parameter is set to
$\kappa=0.1$ and the environment is therefore dominated by the
ferromagnetic interaction.  In the absence of perturbation we have two
almost degenerate eigenvalues, the gap to the third lowest state is
large. For small values of $\Delta$ the overlap between the ground
state $|0\rangle_E$ of $H_E$ and the ground state of the perturbed
enviroment $|0^{\uparrow}\rangle_E$ is very close to one $\langle
0^{\uparrow}|0\rangle_E\approx 1$. At $\Delta\approx 0.2$ there is an avoided
level crossing between the two lowest levels. Close to the avoided 
crossing, the eigenvectors of the two states evolve rapidly in Hilbert
space and end up switching directions.~\cite{PhysRevA.23.3107}
Thus, after the level crossing the first exited state overlaps
completely with what was the the ground state before the level
crossing took place $\langle 1^{\uparrow}|0\rangle\approx 1$.  The overlap with
the the ground state of $H_E$ is, however, still very close to one as
long as only two states take part in the crossing. The eigenvector
corresponding to a large overlap with the original ground state has
simply been swapped with its neighbour and the coherence of the
central system is conserved according to Eq.~\eqref{olapsimp}.

When the disorder of $H_E$ increases, the picture becomes more
complex.  In Fig.~\ref{ordered05}~(middle) we plot the 20 lowest eigenvalues
against $\Delta$, but we use a higher degree of disorder in the
environment ($\kappa=0.5$). Since the environment has a larger
contribution from frustrated couplings in $H_E$, the spacing between
the energy levels is more uniform due to the level repulsion effect.~\cite{mehta}
 In this particular case, the energy of the original
ground state $|0\rangle_E$ is shifted upwards by the perturbation. The
energy of this state can be tracked by the dark line, highlighting the
eigenvalues corresponding to eigenvectors with large overlap element
with the original ground state. In Fig.~\ref{ordered05}~(middle) we can compare
the eigenvalues with the maximal overlap element. We find that the
reduction in the maximal overlap element correspond to values of
$\Delta$ where avoided level crossings take place. For large values of
$\Delta$ the levels are closer, and we find avoided crossings where
three or more levels are involved. Thus the overlap element is split
between several states. The maximal overlap element is therefore
reduced at these values of $\Delta$.

Having developed the sufficient understanding, we are now also able to explain
the oscillatory behavior during the ``phase transition'' in Fig.~\ref{N9P6_D}.
In the region of the transition ($\kappa\approx 0.5$), more levels are present
close to the ground state, and the repulsion width increases with $\kappa$.
When the ground state gets close enough to the first exited state to feel repulsion, the corresponding
eigenvectors begin rotating in the subspace they span. The overlap element is initially reduced
and transferred to the first exited state. Eventually, the first excited state will be the closest
in Hilbert space to the original ground state $\left|0\right\rangle$, explaining the sharp
cusps of Fig.~\ref{N9P6_D}. When $\kappa$ is increased even more, the picture grows more complex
as several levels are involved.

In Fig.~\ref{ordered05}~(bottom)  we have reduced the ferromagnetic part of
$H_E$ to zero ($\kappa=1.0$). In this spin-glass phase the effect of level
repulsion is strongly pronounced. The space between levels at which the
eigenvectors start to repel each other is related to the size of the
off-diagonal elements of the Hamiltonian in the basis of the
perturbation (in this particular case -- the coupling to the central spin
in the $S_i^z$ eigenbasis).~\cite{hsu} When $\kappa$ is large, the
off-diagonal elements in the Hamiltonian, Eq.~\eqref{ham}, are larger
than the average level spacing. This means that avoided crossings take
place continuously as the parameter $\Delta$ is increased. In the
parameter range where the distance between levels is smaller than the
width of repulsion, the eigenvectors will in general evolve with
$\Delta$ in the Hilbert space spanned by the eigenvectors of the
repelling levels.

Thus, we find a crossover between two regimes. In the weak repulsion
regime, the repulsion width is smaller than the typical distance
between levels. In this regime we will have few and pronounced avoided
crossings, the crossings will typically involve only two levels and
the probability of multi-level crossings is strongly suppressed. Each
two level avoided crossing will result in a swap between the
eigenvectors involved, but does not reduce the largest overlap element
after the crossing has taken place. The overlap element is reduced
only during the crossing, still the coherence of the central system is
only slightly altered, due to the levels approaching degeneracy.  In
the second regime, we have strong level repulsion. In this regime the
repulsion width is of the same order or larger than the typical
distance between levels such that each level is for a large range of
$\Delta$ repelled by more than one level at the same time. When the
repulsion width is much larger than the average level splitting, a
large fraction of the levels become connected in the sense that if a
certain level is repelled by the one or more levels above and these
again is repelled by the next few levels that are again repelled by
new levels. The corresponding eigenvectors will then evolve
continuously in the Hilbert space spanned by this cluster of levels.

\begin{figure}[h]
\centerline{
  \includegraphics[width=8cm]{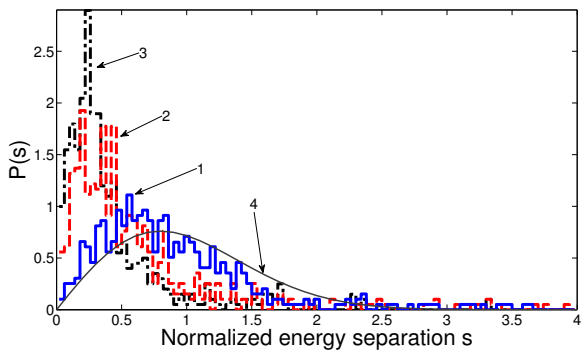}
}
  \caption{(Color online) Level spacing distribution different degrees of disorder. 1 (solid, blue) - spin glass, $\kappa$ =1; 2 (dashed, red) - intermediate frustration,  $\kappa=0.5$; 3 (dashed/dots, black) - ferromagnetic phase, $\kappa=0.25$.
%
For comparison we also plot the Wigner-Dyson
 distribution $P(s)=(s\pi/2)e^{-s^2\pi/4}$ ($4$, thin solid, grey).
The number of spins in the environment is $N=10$.
} 
\label{wigner}
\end{figure}

The energy levels of a system where the repulsion width is larger than
the level splitting is expected to be characterized by a distribution
of energy levels following Wigner-Dyson statistics.~\cite{hsu} In
Fig.~\ref{wigner} we plot the level-spacing distribution of $H_E$ for
different values of the disorder parameter $\kappa$. For large values
of $\kappa$ we find that the distribution is consistent with the
Wigner-Dyson distribution implying that the repulsion width is larger
than the average splitting. 
At the same time, for small values of $\kappa$, 
where we have a ferromagnet, we find  a special
distribution of eigenvalues with two (almost) degenerate ground
states (i.e. $\left|\uparrow\uparrow...\uparrow\right\rangle_E$ and 
$\left|\downarrow\downarrow...\downarrow\right\rangle_E$) and the next levels having a high degree of
degeneracy. Each of the two ground states correspond to the bottom of a potential well,
 excited states belonging to different wells cannot be connected by flipping of two
spins. 
The statistics obtained in Fig.~\ref{wigner} is therefore
sorted by magnetization, the level statistics for each potential well of $H_E$ is treated separately.

In summary, we find a weak repulsion regime, when $H_E$ has a low
degree of disorder. In this regime the overlap element, $\langle
n^{\uparrow}|0\rangle_E$, between the original ground state and the set of
eigenstates of the Hamiltonian in presence of the central spin is
conserved even if we make the coupling to the central spin strong.  In
the second regime, when $H_E$ has high degree of disorder, we have
strong repulsion between large clusters of states. In this regime, the
set of eigenvectors of $H_E$ is very sensitive to the presence of the
central spin.  The largest overlap element, $\langle n^{\uparrow}|0\rangle_E$, is
therefore rapidly reduced as the coupling to the central spin is
increased.

\subsection{The initial state of the environment}
\label{temp}

\begin{figure}[h]
\centerline{
  \includegraphics[width=8cm]{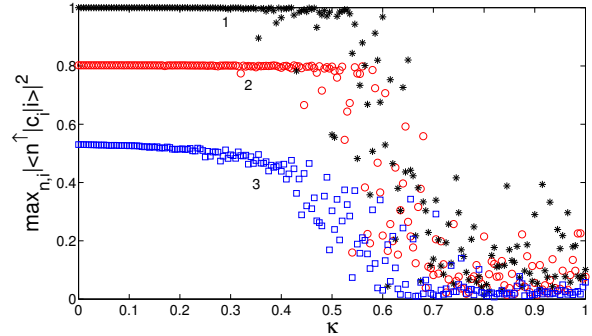}
}
  \caption{The largest overlap element plotted versus the disorder strength $\kappa$: $1.$ low temperature $T=0.01$ (stars, black),
$2.$ intermediate temperature $T=0.10$ (circles, red) and $3.$ high temperature $T=1.00$ (squares, blue). The coupling to the central spin is $\Delta=2.0$ and the number of spins in the environment is $N=9$.}
\label{Tran}
\end{figure}

In Ref.~\onlinecite{yuan_prb08}, the importance of the initial state of the 
environment was studied. More efficient and stable decoherence was found for an initial state corresponding to
infinite temperature, however no detailed explanation of this observation was given.
If the initial state of the environment is no longer the ground state, 
but a linear combination of eigenstates from the set $\{|n\rangle_E\}$ such that
$|\psi_0\rangle_E=\sum\limits_i c_i|i\rangle_E$, where
$|i\rangle_E\in\{|n\rangle_E\}$, 
Eq.~\eqref{olapsimp}  has to be replaced by
\begin{align}
\rho_{\uparrow\downarrow}^S&=\sum\limits_{n,i}|c_i\langle n^{\uparrow}|i\rangle_E|^2 e^{i(E_n'-E_i)t}.
\label{olapsimpT}
\end{align} 
For finite temperature the overlaps are distributed over a
number of eigenstates according to their Boltzmann weight,
$e^{-E/kT}$. The coherence of the central spin, however, is
conserved, $|\rho_{\uparrow\downarrow}^S|=1$, as long as the
perturbation introduced by the central spin does not alter the
eigenvalues of the environment. If there is a significant
perturbation, the coherence is reduced by an 
additional factor given by the square of the largest amplitude of the expansion
$|\psi_0\rangle_E=\sum\limits_i c_i|i\rangle$.  The effect is shown
in Fig.~\ref{Tran}.

\subsection{Enhancement of coherence by an external magnetic field}
\label{magnetic}
\begin{figure}[htb]
\centerline{
  \includegraphics[width=8cm]{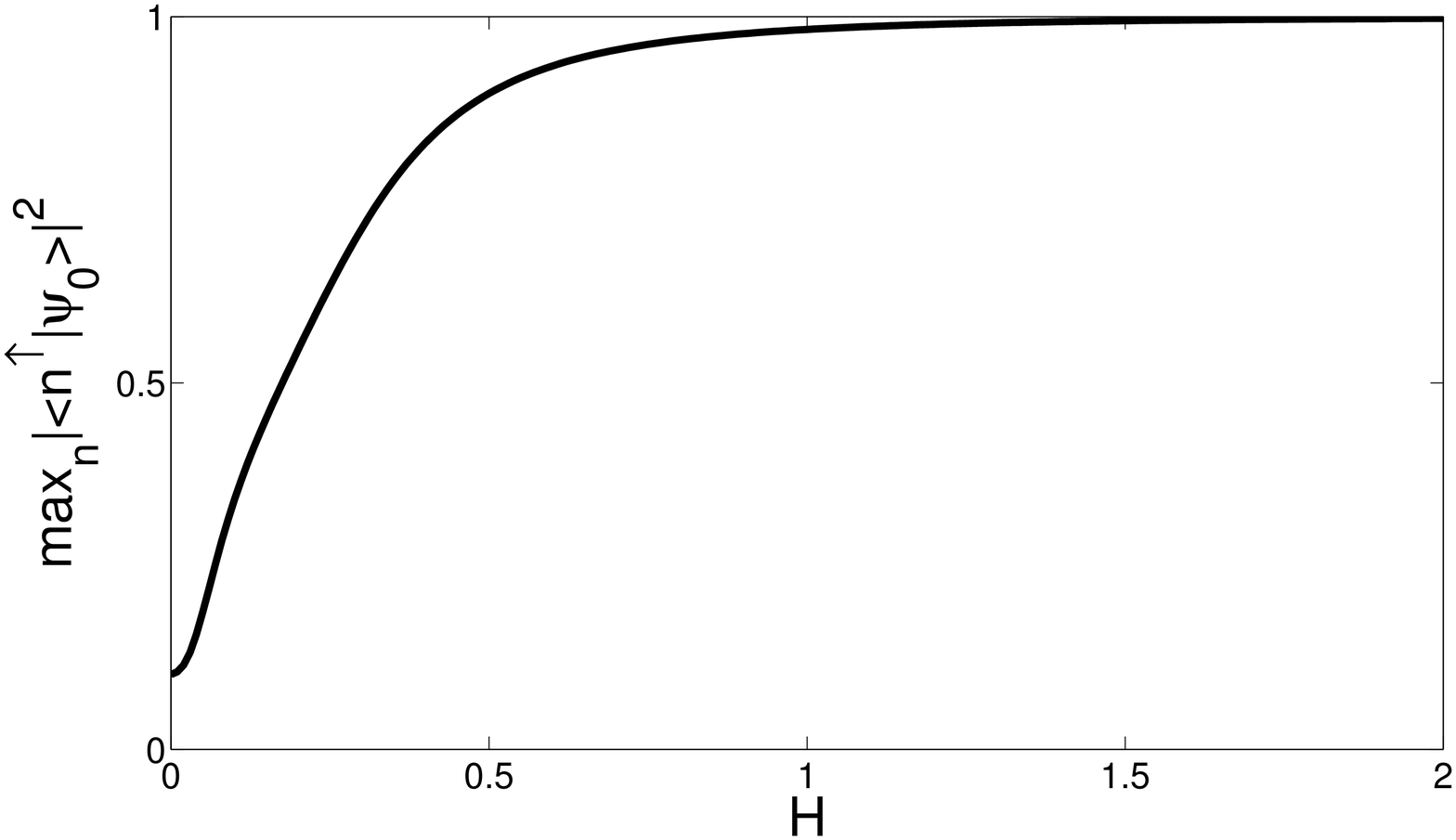}
}
  \caption{The largest overlap element $|\langle n^{\uparrow}|\psi_0\rangle_E|^2$
plotted versus external magnetic field $H$.  The disorder parameter is
$\kappa=1.0$, the coupling to the central spin is $\Delta=\sqrt{N}$
and the number of spins in the environment is $N=9$.}
\label{mag}
\end{figure}
As a concequence of the preceding analysis we find that the presence
of an external magnetic field, $H_{\text{ext}}$, might enhance the coherence
of the central system.  The magnetic field will polarize the spins in
the environemt, resulting in a larger overlap element between the
ground state $|\psi_0\rangle_E$ of the unperturbed environment $H_E$ and
the set of eigenstates $\{|n^{\uparrow}\rangle_E\}$ in presence of the central
spin.  When the magnetic field is sufficiently strong to break the
frustration in the ground state $\left|0\right\rangle_E$, i.~e., when the
magnitude of the external field is of the same order as the coupling
between the spins in the environment, $H_{ext}\gg \sqrt{N}\Omega$, the
presence of the central spin will not significantly alter the
magnetized ground state of the environment unless the coupling to the
environment is strong compared to the external field.  Thus, if the
spin environment of the central system is disordered, magnetization is
beneficial to the coherence of the central system. This procedure has
already been applied experimentaly, see
Refs.~\onlinecite{Hanson_science,takahashi}.

\section{Discussion}
\label{discussion}
In this article, we have considered the special case where the central
spin is coupled diagonally to its environment. Then no transitions can
take place between the eigenstates of the central system and the
decoherence is entirely due to renormalization of its energy splitting
(pure dephasing). This choice of coupling simplifies the treatment
since the effect of the central spin upon the environment can be
treated as a static perturbation.  If we loosen this restriction and
also include real transitions between the eigenstates ($T_1$-processes)
 the central system will participate in the complex
many-body dynamics of the total system. However, if the number of
spins in the environment is large, the fine details of the coupling,
$H_{SE}$, should not result in qualitatively different behavior of
the environment with respect to level repulsion. The microscopic
details of the dynamics will, of course, strongly depend on the exact
nature of the coupling. We, therefore, believe that the central spin
will preserve its coherence much longer in the ordered environment,
compared to a frustrated environment also in the presence of
non-diagonal system-environment coupling $H_{SE}$. The numerical
analysis in Refs.~\onlinecite{yuan_prb08,yuan_pra07} support this
hypothesis.

We considered an arrangement where the central system coupled to each
spin in the environment. In the presence of a very large environment,
where the connectivity between the subsystems is limited, this
approximation might fail. As an example the central spin might couple
to only a few spins of the environment. However, even if the central
spin couple only to few spins, in the presence of a ferromagnetic
environment, this might be sufficient for coupling to collective modes
of excitation (i.~e., spin waves).

Since we treat a closed quantum system, we do not expect details of
our analysis to carry on to realistic open systems. In the
thermodynamic limit we expect that the environment will be damped,
forgetting interactions with the central spin at times earlier than
the correlation time. However, the analysis should be relevant to
systems where the effective temperature is much less than the typical
splitting between states in the environment.

We found it useful, in light of the correlations shown in Fig.~\ref{corrdec}, 
to discuss the decoherence of the central spin in terms of
the overlap elements between the ground state of $H_E$ and the eigenstates 
$\{\left|n^{\uparrow}\right\rangle\}$ of the environment in presence of the central spin.
However, the largest overlap element of Eq.~\eqref{olapsimp} does not necessarily give the whole picture.
The coherence of the central spin may differ from what was predicted by the 
overlap element due to the phase factor $e^{i(E_n^{\uparrow}-E_m^{\downarrow})t}$. 
If the ground state of $H_E$ is degenerate due to symmetry, 
and the central system is unable to break this symmetry, 
then coherence will persist in the central system even if the overlap with 
the ground state of $H_E$ is split between several degenerate states. 
If the degeneracy is not exact, coherence might still decay extremely slowly 
if the difference $\left|E_n^{\uparrow}-E_m^{\downarrow}\right|$ of the states overlapping 
with $\left|\psi_0\right\rangle_E$ is small.

\section{Conclusion}
\label{conclusion}

In conclusion, we have analyzed the efficiency of decoherence 
using the overlap elements $\langle n^{\uparrow}|0\rangle_E$, between the
ground state of the isolated environment and the set of eigenstates of
the environment in presence of the central spin.  It was shown that
the square of the largest overlap element, $\max_n|\left\langle n^{\uparrow}|\psi_0\right\rangle|^2$,
is a very good indicator for the efficiency of decoherence.
The size of the largest overlap element tends to be much larger for an
environment with no competing interactions, than in case of a
environment with many frustrated coupling. The undelying mechanism
behind this effect can be explained by the statistics of the
eigenvalues of $H_E$. Coupling to an external object, e.~g., a central
spin, results in avoided level crossings between the levels of the
environment. In the absence of frustration, the level repulsion is
weak and the avoided crossings will take place in a short interval in
the coupling parameter to the external object, $\Delta$. The
eigenvectors corresponding to the involved levels will simply switch,
and the overlap element remains unaltered. In this weak repulsion
regime, multi-level crossings are strongly suppressed.
In the opposite regime, characterized by strong level repulsion,
eigenvalues within large fractions of Hilbert space are subject to
mutual level repulsion. In this strong repulsion regime the
corresponding eigenvectors will rapidly mix when increasing $\Delta$
resulting in very efficient decoherence of the central object.  

As a side note -- we have shown that a external magnetic field can transfer the
environment from the strong to the weak repulsion regime provided it
is stronger than the frustrated couplings present, thereby enhancing
the coherence of the central spin. Thus, it should be possible to
enhance the coherence time of a central spin in the presence of a
spin-glass like environment, by applying an external magnetizing field
that is of the same magnitude or larger than the internal coupling in
the environment.

\appendix

\end{document}